\documentclass[preprint,12pt]{elsarticle}

\usepackage{amssymb}
\makeatletter
\def\ps@pprintTitle{%
  \let\@oddhead\@empty
  \let\@evenhead\@empty
  \let\@oddfoot\@empty
  \let\@evenfoot\@oddfoot
}
\makeatother

\journal{}

\begin{document}

\begin{frontmatter}


\author{Tamio Yamazaki\corref{cor1}}
\ead{yamazaki.tamio@canon.co.jp}
\cortext[cor1]{Corresponding author. Tel.: +81-3-3758-2111.}

\title{Parametrization of coarse grained force fields for dynamic property of ethylene glycol oligomers/water binary mixtures}


%%\author{}

\address{Analysis technology center, Canon Inc. 3-30-20, Shimomaruko,
 Ota-ku, Tokyo 146-8501, Japan.}

\begin{abstract}
 To evaluate shear viscosity of ethylene glycol oligomers (EGO)/water binary mixture by
 means of coarse-grained molecular dynamics (CG-MD) simulations, we proposed the self-diffusion-coefficient-based parameterization of non-bonded interactions among CG particles.
 Our parameterization procedure consists of three steps: 1$)$ determination of bonded potentials,
 2$)$ scaling for time and solvent diffusivity , and 3$)$ optimization of Lennard-Jones parameters
 to reproduce experimental self-diffusion coefficient and density data.
 With the determined parameters and the scaling relations, we evaluated shear viscosities of
 aqueous solutions of EGOs with various molecular weights and concentrations. Our simulation result
 are in close agreement with the experimental data. The largest simulation in this article corresponds to a 1.2 $\mu$s atomistic simulation for 100,000 atoms. Our CG model with the parameterization scheme for CG particles may be useful to study the dynamic properties of a liquid which contains relatively low molecular weight polymers or oligomers.  
\end{abstract}


\end{frontmatter}

% \linenumbers

\newpage
\section{Introduction}
\hspace*{1em}Aqueous polymer solutions are widely used in industrial and household
 applications. For example, in ink-jet printing, polymer additives are used to
 control fixity and osmosis of the ink droplets to the paper.
 In order to estimate such properties, it is very important to understand the shear
 viscosity and the diffusivity of the polymer solutions.
 The molecular dynamics (MD) method at atomic level is the most common technique to
 estimate the shear viscosity of a solution or the self-diffusion coefficient of
 a component \cite{Allen1989,Frenkel2001}.
 However, even if a small oligomer of which the number of monomers is around 10,
 the longest relaxation time (for usual polymer solutions, it will be the rotational
 relaxation time of polymer chains), is several tens nanoseconds, which is about
 1000 times longer than the characteristic time of water molecules.
 The rotational relaxation time of the polymer chains increases in proportion to
 some power of its molecular weight. 
\\
\hspace*{1em}To obtain the ensemble averaged quantities, the sampling time
 (the time-average period) should be longer than the longest relaxation time
 of the system.
 The extreme long relaxation time complicates the estimations of the shear
 viscosity and the self-diffusion coefficient by using all-atom
 molecular dynamics (AA-MD) of which typical time-steps is femto-second order. 
\\
\hspace*{1em}In order to extend the accessible time scale, the coarse-grained
 (CG) techniques have been developed for several decades. The CG techniques,
 in which some atomic groups are represented by a single particle, are widely
 used for simulations of meso-scale phenomena in soft materials (lipids,
 surfactants and polymers), for example,
 vesicle formation and fusion \cite{Vesicle2006,Vesicles2003},
 self-assembly \cite{Copolymer2004,PCCPHatakeyama2007},
 micelle formation \cite{Shinoda2007,Micelle2007,Micelle_CGMD2002,
surface_oilwater1990} and
 pore formation in lipid bilayer \cite{Pore2006,Pore2006B,Pore2004,Pore2008}.
 The coarse grained molecular dynamics (CG-MD) can achieve speed-up of up to
 several orders of magnitude faster than an AA-MD. Of course, the degree of
 speed-up depends on the details of the CG model or the system size.
 CG-MD method is very powerful tool to study of the static property, such as
 equilibrium structure of a large system, and it is also successful in
 investigating the fundamental mechanism of the long time-scale behaviors of
 soft materials. 
\\
\hspace*{1em}To apply this method to a quantitative estimation of the dynamics,
 it should be noted that the time in CG-MD trajectory is not equivalent to the
 time in the all-atom (AA) description, due to the lack of atomistic details
 in the CG model. An effective method to obtain the consistency of the molecular
 motions between AA-MD and CG-MD is to multiply the time scale of the CG-MD by
 a constant factor.
\\
\hspace*{1em}Recently, several authors reported the systematic studies of the
 dynamical and rheological properties of polymer systems polymer systems(polystyrene
 melt \cite{Kremer2009, Carbone2008}, polyamide-6,6 melt \cite{Carbone2008} and
 aqueous polyethylene glycol solutions \cite{Fischer2008}) by the multiscale approach
 that combines AA-MD and CG-MD simulations through the scaling of time in the CG model.
 With the well-tuned potential energy function among the particles in the
 CG model, CG-MD gives us the reasonable possibility to investigate both
 statical and dynamical properties of large scale systems which include macromolecules.
 However, there are few studies about the parametrization for the non-bonded CG
 interactions, which can be adapted to the estimations of dynamic properties
 of system, especially of the multi-component system.
\\
\hspace*{1em}In this article, we will present the results of the CG-MD
 simulations, including the systematic determination of the non-bonded
 parameters for the ethylene glycol oligomer (EGO)/water binary mixtures
 based on the self-diffusion coefficient and density data. We will also
 present the results of the shear viscosity of the mixture estimated
 by means of the nonequilibrium CG-MD simulations with the newly
 determined force field parameters.
\\
 The article is organized as follows. The experimental methods for
 measurement of the shear viscosity and self-diffusion coefficient
 are explained in the section two. The explanation of the CG model
 and the force-field parameterization for EGO/water binary mixtures
 and the computational details are in the 3rd section. The results
 of the comparison between our simulations and the experimental
 measurements are shown in the Results and discussion section
 that follows.

\section{Experimental section}

\subsection{Measurement of the shear viscosity and the self-diffusion coefficient}
\subsubsection{Materials   }
 Reagent grade diethylene glycol (DEG), tetraethylene glycol (TEG) and PEG600 were
 purchased from Sigma Aldrich Chemical Company and used in this work without further
 purification. EGO/water binary mixtures were prepared gravimetrically using distilled
 water. D$_{\rm{2}}$O with a minimum deuteration degree of 99.95 \% (Merck Co. \& Inc.,
 Darmstadt, Germany) was used for all experiments as the NMR lock solvent.

\subsubsection{Experimental details   }
 The shear viscosities of the EGO aqueous solutions were measured at 293 K by
 the RE80 viscometer manufactured by Toki Sangyo Co., Ltd. 
\\
 The self-diffusion coefficient measurements were performed at 293 K by pulse field
 gradient spin echo (PGSE)\cite{PGSE1987} using standard $ledbpgp2s$ sequence on the
 Avance600 NMR spectrometer (Bruker BioSpin GmbH, Rheinstetten, Germany). To avoid mixing
 H$_{\rm{2}}$O in the aqueous EGO solution and D$_{\rm{2}}$O as the NMR lock solvent, we
 used NMR tubes which have a double tube structure. In the NMR tube, the inner tube was
 filled with the NMR lock solvent D$_{\rm{2}}$O, and the outer tube was filled with
 aqueous EGO solution. 
\\
 In $^{1}$H-NMR spectra of aqueous EGO solutions, the peaks of around $\delta \approx3.6$
 ppm are assigned to the protons of CH$_{\rm{2}}$CH$_{\rm{2}}$O groups of EGO, and
 the peaks of around $\delta \approx4.7$ ppm are assigned to the protons of water
 molecule. The self-diffusion coefficients of assigned peaks are abbreviated in
 this article as $D_{\rm{EGO}}$ ($\delta \approx3.6$ ppm) and $D_{\rm{OH}}$
($\delta \approx4.7$ ppm). Due to the proton exchange between hydroxyl groups of
 EGO end and water molecules, the diffusivity of EGO have a strong influence on
 the $D_{\rm{OH}}$.
\\
 Under the assumption that the proton exchange between hydroxyl of EGO end and water molecules
 is very rapid, a self-diffusion coefficient of water ($D_{\rm{w}}$) can be obtained by
\cite{Vergara1999}
\begin{equation}
D_{\rm{w}} = \frac{\chi_{i}}{1-\chi_{i}}D_{\rm{OH}} - \frac{1}{1-\chi_{i}}D_{\rm{EGO}} ,  
\label{eq:(1)}
\end{equation}
where $\chi_{i}$ is the molar fraction of hydroxylic protons of EGO ends, and $D_{\rm{w}}$
 is the self-diffusion coefficient of water molecule.

\section{Theoretical section}

\subsection{Coarse-graining of ethylene glycol oligomer and water }
In our coarse-grained model, EGO molecules and water molecules are represented by coarse-grained particles,
 as shown in Figure 1.
\begin{figure}[h]
\centering
  \includegraphics[height=2cm]{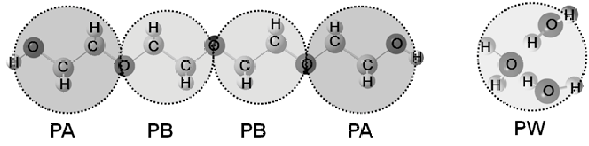}
  \caption{Coarse-graining scheme of a tetraethylene glycol and water. Water is modeled as a PW particle
 which corresponds to three water molecules.}
  \label{fgr:example}
\end{figure}

 EGO molecules are modeled by two types of particles ("PA") and ("PB"). The PA particle
 represents the oxyethylene unit of both ends of a ethylene glycol chain, and the PB
 particle represents the oxyethylene unit in a ethylene glycol chain.
 The mass of PA and PB particles are 53 amu and 44 amu respectively.
 A mass of an oxygen atom of ether is distributed to two adjoined coarse grained particles
 at an equal ratio.
\\
 Water is modeled by a single particle ("PW") corresponding to three real water molecules.
 The mass of PW particle is 54 amu. DEG, TEG and PEG600 molecules are modeled as PA-PA,
 PA-(PB)$_{\rm{2}}$-PA and PA-(PB)$_{\rm{11}}$-PA, and we expressed these EGO species as EGO2,
 EGO4 and EGO13 respectively.

\subsection{Coarse-grained pair potential}
We assume that the total potential energy for CG molecule is written as
\begin{equation}
U_{\rm{tot}} = \sum U_{\rm{b}}(L_{ij}) + \sum U_{\rm{ang}}(\Theta_{ijk}) + \sum U_{\rm{non}}(R_{ij}), 
\end{equation}
where the terms, $U_{\rm{b}}$, $U_{\rm{ang}}$, $U_{\rm{non}}$ are the effective
 potential functions of bond length $L_{ij}$, bond angle $\Theta_{ijk}$, and the distance
 between non-bonded CG particles $R_{ij}$ respectively.
 Here, $i$, $j$ and $k$ are the indices of the CG particles.
 The non-bonded CG interactions $U_{\rm{non}}(R_{ij})$ are modeled using
 the 12-6 Lenard-Jones potential, 

\begin{equation}
 U_{\rm{non}}(R_{ij})=4 \epsilon_{ij} \left(\left(\frac{\sigma_{ij}}{R_{ij}}\right)^{12}
        -\left(\frac{\sigma_{ij}}{R_{ij}}\right)^{6}\right),
\end{equation}
where $\sigma_{ij}$ is the (finite) distance at which the inter-particle potential is zero,
 $\epsilon_{ij}$ is the depth of the
 potential well. We use the Lorentz-Berthelot combination rule for interaction between the
 different species,
\begin{equation}
\epsilon_{ij} = \sqrt{\epsilon_i \cdot \epsilon_j}\hspace{1mm},\hspace{10mm}     
\sigma_{ij} = \frac{\sigma_i + \sigma_j}{2} . 
\end{equation}
In the case the interaction between PB and PW particles,$\hspace{1mm}{\epsilon}_{ij}$ 
is calculated from the eq.(5) instead of eq.(4), 
\begin{equation}
\epsilon_{ij}= \gamma\cdot\sqrt{\epsilon_i \cdot \epsilon_j}\hspace{1mm}.
\end{equation}

\subsection{Scaling relations  }
 As mentioned in Introduction, whenever the CG model is used, due to the lack of atomistic details,
 the speed of the time-evolution of the CG-MD trajectory is inconsistent with that of the AA-MD
 trajectory. To estimate of dynamical properties of the aqueous EGO solutions by means of CG-MD,
 we introduced the time mapping parameter $S$, which is defined by Kremer et.al.\cite{Kremer2009}
 as a ratio between the effective segment frictions in the AA model and in the CG model,
\begin{equation}
S\equiv\frac{\zeta^{\rm{AA}}}{\zeta^{\rm{CG}}} , 
\end{equation}
 where $\zeta^{\rm{AA}}$ is an effective scalar friction coefficient of a segment in the AA model,
 and as $\zeta^{\rm{CG}}$ is ones in the CG model.
 In this article, we have assumed that the ratio between the shear viscosities in the AA model
 ${\eta^{\rm{AA}}}$ and in the CG model ${\eta^{\rm{CG}}}$ is equal to the parameter $S$,
 namely: 
\begin{equation}
\eta^{\rm{AA}} = {S}\cdot{\eta^{\rm{CG}}} .  
\end{equation}
 According to the Stokes-Einstein relation, the self-diffusion coefficient $D$ is
 inversely proportional to $\eta$ and the hydrodynamic radius of a segment $r_{\rm{h}}$,
\begin{equation} 
D\propto(\eta\cdot r_{\rm{h}})^{-1} .
\end{equation}
 In our CG model, three water molecules are included in a single PW particle. The PW particle in
 the CG model has a larger hydrodynamic radius than the individual realistic water molecule,
 and thus the self-diffusion coefficient of water molecule in CG model is somewhat smaller than
 that in the AA model, even if we apply the time mapping parameter $S$ properly.
\\
 We define the hydrodynamic radius ${r_{\rm{h}}^{\rm{CG}}}$ of the molecule in CG model by
\begin{equation}
r_{\rm{h}}^{\rm{CG}}=\sqrt[3]{n}\cdot{r_{\rm{h}}^{\rm{AA}}} ,
\end{equation}
 where $n$ is a number of atomistic molecules, which are included in one coarse-grained molecule,
 $n$ = 1 (for EGO), $n$ = 3 (for water) as shown in Figure 1, and ${r_{\rm{h}}^{\rm{AA}}}$ is the
 hydrodynamic radius of molecule in the AA model, the factor $\sqrt[3]{n}$ in eq.(9) comes from
 the assumption of the linear relationship between the cube of the hydrodynamic radius of a 
molecule and its volume.
 From eqs.(7) and (8), the scaling relation between self-diffusion coefficients of molecular systems,
 which are described by the AA model and by the CG model, is given by 
\begin{equation}
D^{\rm{AA}} = {S^{-1}}\cdot{\sqrt[3]{n}}\cdot{D^{\rm{CG}}} ,
\end{equation}
 where $D^{\rm{CG}}$ is defined using the Einstein relation,
\begin{equation}
D^{\rm{CG}} = \lim_{{t} \to \infty}{\frac{1}{6t}}
\left<\left({r^{\rm{CG}}_{\rm{com}}}(t)-{r^{\rm{CG}}_{\rm{com}}}(0)\right)^{2}\right> ,    
\end{equation}
 where $r^{\rm{CG}}_{\rm{com}}(t)$ is coordinates vector of the center of mass of the molecule in CG model
 at time $t$, $\left<\cdots\right>$ denotes the ensemble average. In this article, $S$ is dealt
 with as a constant value for simplification, though the large molecular weight dependence of $S$
 is argued in the CG-MD study of polystylene melt in Ref.15. 
\\
 In order to determine $S$, we consider the self-diffusion coefficient of pure water using CG-MD
 simulation $D^{\rm{CG}}_{\rm{water}}$, and the experimental value $D^{\rm{EXP}}_{\rm{water}}$
 observed by NMR measurement. From eq.(10), $S$ is given by 
\begin{equation}
S = \frac{\sqrt[3]{3}\cdot{D^{\rm{CG}}_{\rm{water}}}}{D^{\rm{EXP}}_{\rm{water}}} . 
\end{equation}

\subsection{Parameterization of bonded potentials of EGO chain}
 The Boltzmann inversion method is well-known as one of the techniques to evaluate effective
 mesoscale potential from atomistic simulation \cite{KremerBI1998}. This technique enables us
 to determine the bonded intramolecular interaction especially.
 In this article, $U_{\rm{b}}(L_{ij})$ and $U_{\rm{ang}}(\Theta_{ijk})$ are determined by this
 method. In a thermal equilibrium system, we assume that an appearance probability $P$ of a
 state vector of a system $Q$, obeys the Boltzmann distribution.
\begin{equation} 
P(Q)\propto{{\rm{exp}}(-\beta{U}(Q))} , 
\end{equation} 
where $U(Q)$ is a certain effective potential energy as a function of $Q$ , and $\beta$ is
 $1/k_{\rm{B}}T$ .
\\
Once $P(Q)$ is obtained, a effective potential $U(Q)$ can be straightforwardly determined
 from the inversion of eq.(13).
 In this article, we assumed that all of the bonded (stretching / bending) potentials in the
 EGO chain are approximately given by the same $U$, as we will show in eq.(15).  
\\
 To obtain $P(L_{ij})$ and $P(\Theta_{ijk})$, the AA-MD simulation of a triethylene glycol
 dimethyl ether (TEGDE) in gas-phase is performed at 293 K. With our coarse-graining manner,
 a TEGDE molecule is modeled by three coarse-grained particles $({{\rm{PB}}_{p}}, {p}=1,2,3)$
 as shown in Figure 2.
\begin{figure}[h]
\centering
  \includegraphics[height=2cm]{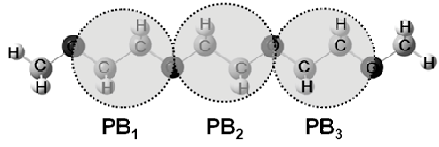}
  \caption{Coarse-graining scheme of a triethylene glycol dimethyl ether molecule.}
  \label{fgr:example}
\end{figure}

The center of PB particle is defined as the center of mass of the oxyethylene monomer unit,
 in which two oxygen atoms at both ends of the unit are weighted by 0.5 and other atoms are
 weighted by 1.0. Histogram $H(L_{ij})$ of bond length (bond ${\rm PB_{1}}-{\rm PB_{2}}$ and
 bond ${\rm PB_{2}}-{\rm PB_{3}}$) and histogram $H(\Theta_{ijk})$ of bond angle(bond angle
 ${\rm PB_{1}}-{\rm PB_{2}}-{\rm PB_{3}}$) are obtained from the 50-ns trajectory of the
 atomistic molecular dynamics simulation. Then, these histograms are renormalized by 
\begin{equation} 
  P(L_{ij})\propto\frac{H(L_{ij})}{{L_{ij}}^{2}} \hspace{1mm},\hspace{10mm}  
  P(\Theta_{ijk})\propto\frac{H(\Theta_{ijk})}{{\rm{sin}}(\Theta_{ijk})} .   
\end{equation} 
 As a technical subject for the $U(Q)$, because there are a lot of noises in the potential
 energy function obtained by the Boltzmann inversion scheme, it should be smooth by some
 appropriate functions.
 We assumed that the probability distribution function $P(Q)$ can be expressed by the
 linear combination of Gauss functions $G_{l}$ \cite{MCBI2005}. Then the effective potential
 energy functions $U(L_{ij})$ and $U(\Theta_{ijk})$ are obtained by
\begin{eqnarray} 
U(Q)/{k_{\rm{B}}T}&\hspace{-2mm}=&\hspace{-2mm}-\ln(P(Q)) 
=-\ln(\sum_{l=1}^{m}{G_l}) + const. \nonumber \\
&\hspace{-3mm}=&\hspace{-2mm}-\ln\sum_{l=1}^{m}
 \frac{A_{l}}{\xi_{l}\sqrt{\pi/2}}\exp\left(-\frac{(Q-\mu_l)^2} {2{\xi_l}^{2}}\right)
 + const.\hspace{1mm}, \nonumber \\ 
\end{eqnarray}
where $A_l$, $\mu_l$ and $\xi_l$ are total area, center position and width of the Gauss
 function $G_l$ respectively, and $m$ is the number of the Gauss functions for smoothing
 of $P(Q)$. In this article, we decided $m$ = 3 for the both potentials of bond length
 $(Q=L_{ij})$ and bond angle $(Q=\Theta_{ijk})$. The areas ($A_l$), center coordinates
 ($\mu_l$) and widths ($\xi_l$) are determined by curve fitting of the calculated bond
 length / bond angle distribution data.
\\
\subsection{Parameterization of Non-bonded potentials}
In our CG model for the EGO / water binary system, there are three different particles (PA, PB, PW), and 7 parameters for the non-bonded interaction $U_{\rm{non}}(R_{ij})$, which are $\epsilon_{\rm{PW}}$ $\epsilon_{\rm{PA}}$, $\epsilon_{\rm{PW}}$, $\sigma_{\rm{PW}}$, $\sigma_{\rm{PA}}$,
 $\sigma_{\rm{PB}}$ and $\gamma$. These non-bonded parameters for the $U_{\rm{non}}(R_{ij})$ are determined based on the experimental data (the densities and the self-diffusion coefficients for several EGO / water binary systems). The parameterization procedure consists of the four steps listed below.
\\
STEP 1 : The parameter $\epsilon_{\rm{PW}}$ as a function of parameter $\sigma_{\rm{PW}}$ is determined so as to reproduce the density of pure water at 293 K, which is 0.998 ${\rm g/cm^3}$ \cite{DenWater}. In order to satisfy the experimental pure water density, $\epsilon_{\rm{PW}}$ is uniquely determined according to $\sigma_{\rm{PW}}$. The self-diffusion coefficient of pure water
 $D^{\rm{CG}}_{\rm{water}}$ is evaluated by CG-MD simulation with the fixed parameter values
 ($\epsilon_{\rm{PW}}$ and $\sigma_{\rm{PW}}$). The time mapping parameter $S$ is calculated
 from eq.(12).
\\
STEP 2 : The parameter $\epsilon_{\rm{PA}}$ as a function of parameter $\sigma_{\rm{PA}}$ is determined
 so as to reproduce the density of pure diethylene glycol (EGO2) liquid at 293 K, which is 1.118
 ${\rm g/cm^3}$ \cite{DenPEG}. With the parameters for the PW determined in STEP 1, unique
 $\sigma_{\rm{PA}}$ is obtained through minimizing of the error between the $D^{\rm{CG}}$ calculated
 by CG-MD and the $D^{\rm{AA}}$ observed by NMR of the components in the EGO2/water binary mixtures
 (EGO2 weight fraction : 0.2, 0.5 and 0.8).
\\
STEP 3 : For determination of the parameter $\epsilon_{\rm{PB}}$, we used the density of pure
 tetraethylene glycol (EGO4) liquid at 293 K, which is 1.125 ${\rm g/cm^3}$ \cite{DenPEG}
 and the self-diffusion coefficients of the components in the triethylene glycol (EGO3)/water
 binary mixtures (EGO3 weight fraction : 0.2, 0.5 and 0.8) observed by NMR.
 The EGO4 and EGO3 are modeled in this article as PA-PB-PB-PA and PA-PB-PA, respectively.
 With the parameters for the PW and the PA determined already in STEP 1 and STEP 2,
 the $\sigma_{\rm{PB}}$ can be obtained by the same process in STEP 2. At the first execution
 of STEP 3, $\gamma$ of eq.(5) is setted to 1. After performing STEP 4, $\gamma$ is revised to
 the optimized value. 
\\
STEP 4: In the case of the calculating of the interaction between PB and PW particle, the parameter
 $\gamma$, which adjust the miscibility of the EGO chain in water, should be determined.
 The $\gamma$ is obtained through minimizing of the error between the $D^{\rm{CG}}$ calculated
 by CG-MD and the $D^{\rm{AA}}$ observed by NMR of the components in the PEG600 (EGO13)/water
 binary mixtures (EGO13 weight fraction : 0.2).
\\
\\
STEP 3 and 4 are sequentially repeated until the proper $\sigma_{\rm{PB}}$ and $\gamma$ are obtained.
 In STEP 2, 3 and 4, the root means of square errors (RMSE) are evaluated by 
\begin{equation} 
RMSE=
\sqrt{\frac{1}{M}{\sum_{q=1}^{M}}\left(\log D^{\rm{EXP}}_q - \log D^{\rm{AA}}_q\right)^2} ,
\end{equation} 
where $M$ is the number of data, $D^{\rm{EXP}}_q$ is experimental
 self-diffusion coefficients, $D^{\rm{AA}}_q$ is calculated self-diffusion
 coefficients from eq.(10).

\subsection{Simulation details}
All simulations of this work are performed by GROMACS 4.0.5 \cite{GROMACS4}.
 In the atomistic molecular dynamics simulation of gas phase of TEGDE, the temperature
 is held at $T$ = 293 K by Langevin thermostat \cite{Langevin1976} with correlation time
 $\tau$ = 1.0 ps . The production time step for integration is dt = 1 fs. And cutoff radius
 for LJ and Coulomb potentials is 1.4 nm. The general Amber force field (GAFF) \cite{GAFF2004}
 is used as the atomistic force fields for TEGDE molecule. Atomic charges of a TEGDE molecule
 are assigned by AM1-BCC \cite{AM1_BCC2002} method.
 These force fields and atomic charges are generated by antechamber 1.4 \cite{Antechamber,GAFF2004}
 and acpype 1.0 \cite{ACPYPE}. In the CG-MD simulations, to evaluate the density 
and the equilibrated liquid structure, 2 ns MD simulations are performed in the $NpT$ ensemble.
 Nose-Hoover thermostat \cite{Nose_Hoover1,Nose_Hoover2,Nose_Hoover3} is used at 293 K to control
 temperature of the system. Parrinello-Rahman barostat \cite{NPT_Parrinello1980} is used at 1 atm
 to control pressure of the system.
 In the 2 ns MD simulation, the instantaneous densities are calculated from the last 1 ns trajectory
 every 1000 steps and then are averaged.
 Omitting the first 1 ns data of trajectory as relaxation time, the self-diffusion coefficient is
 calculated from 3 ns (for water, EGO2/water, EGO4/water binary mixture) or 30 ns (for EGO13/water
 binary mixture) MD simulation, which is performed at constant volume $NVT$ ensemble, at the initial structure is the last record of the former 2 ns MD simulation.
\\
 Non-equilibrium molecular dynamics(NEMD) \cite{Hess2002} simulation is applied for the calculation
 of the shear viscosity of aqueous EGO solution. In order to calculate the shear viscosity, 20 ns (for water,
 EGO2/water, EGO4/water binary mixture) or 200 ns (for EGO13/water binary mixture) MD simulation
 is performed. After dropped first 5 ns trajectory, the shear viscosity is calculated by analysis of 
NEMD trajectory.
\\
 There are 5 sets of MD/NEMD simulations with different initial structures and initial velocity
 profiles. The diffusion coefficients or the shear viscosities are calculated from each MD/NEMD
 simulations, and then are averaged. The cutoff radius $R_{\rm{cut}}$ = 1.4 nm for non-bonded
 interaction among the coarse-grained particles and a production time step dt = 10 fs for
 integration of Newton's equation are used as common conditions in  all CG-MD simulations.
\section{Results and discussion }
\subsection{Bonded interactions for PEG chain}
Parameters ($A_l$, $\mu_l$ and $\xi_l$) of eq.(15), which are determined from curve fitting of
 the data of $U(L_{ij})$ and $U(\Theta_{ijk})$ are summarized in Table 1 and Table 2, respectively.

\begin{table}[h]\centering
\small
  \caption{\ Parameters of bond length potential represented by eq.(15)}
  \label{tbl:example}
  \begin{tabular*}{0.9\textwidth}{@{\extracolsep{\fill}}ccccc}
    \hline
    bond type & l & $A_{l}$ & $\mu_{l}$(nm) & $\xi_{l}$(nm) \\
    \hline
   PA-PA,PA-PB and PB-PB & 1 & 0.382 & 0.023 & 0.305 \\
   $  $                  & 2 & 0.229 & 0.020 & 0.338 \\
   $  $                  & 3 & 0.028 & 0.018 & 0.266 \\
    \hline
  \end{tabular*}
\end{table}

\begin{table}[h]\centering
\small
  \caption{\ Parameters of bond angle potential represented by eq.(15)}
  \label{tbl:example}
  \begin{tabular*}{0.9\textwidth}{@{\extracolsep{\fill}}ccccc}
    \hline
    angle type & l & $A_{l}$  & $\mu_{l}(^{\circ})$ & $\xi_{l}(^{\circ})$ \\
    \hline
    PA-PB-PB and PB-PB-PB & 1 & 238.840 & 57.471 & 190.567 \\
    $  $                  & 2 &  45.375 & 24.819 & 123.986 \\
    $  $                  & 3 &  31.826 & 14.765 & 101.560 \\
    \hline
  \end{tabular*}
\end{table}

Figure 3 shows the comparison between the AA model and the CG model for the histograms of
 the length $L_{ij}$ (a) and of the angle $\Theta_{ijk}$ (b) of bonded PB particles of
 TEDME, as shown in Figure 2.
\begin{figure}[!h]
\centering
  \includegraphics[height=11cm]{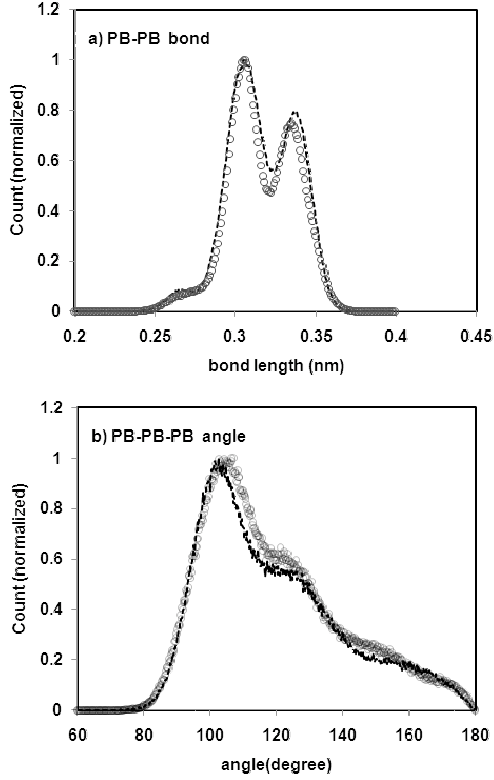}
  \caption{Probability distributions of bond length between bonded PB particles (a)
 and of bond angle among bonded $\rm{PB}_{1}$-$\rm{PB}_{2}$-$\rm{PB}_{3}$ particles
 (b) of TEDME. Open Symbols denote the distribution obtained by using AA-MD simulation.
 Broken line denotes the distribution obtained by CG-MD simulation.}
  \label{fgr:example}
\end{figure}
\subsection{Non-bonded interactions}
The Lennard-Jones parameters for PW, PA and PB particles, the parameter $\gamma$ of
 eq.(5) and the time mapping parameter $S$ are straightforwardly determined by using the
 procedures (STEP 1 to STEP 4) mentioned as section 3.5. In STEP 1, to satisfy the
 experimental liquid density of pure water (0.998 g/$\rm{cm}^{3}$) at 293 K and 1 atm pressure,
 $\epsilon_{\rm{PW}}$ is uniquely determined according to $\sigma_{\rm{PW}}$, which is
 limited in range from 0.38 to 0.42 nm. If the parameter $\sigma_{\rm{PW}}$ is within this
 range, then the PW fluid has a stable liquid phase. Though the $\sigma_{\rm{PW}}$ can be
 selected arbitrary within this range, we selected $\sigma_{\rm{PW}}$ = 0.40 nm. The observed
 self-diffusion coefficient of pure water at 273 K and 1 atm pressure is
 $D^{\rm{EXP}}_{\rm{water}}$ = 2.0 $\rm{m}^{2}$/s by using NMR. From eq.(12), $S$ = 6.19 is
 obtained from eq.(12). Through STEP 2 to STEP 4, the $\sigma_{\rm{PA}}$, $\sigma_{\rm{PB}}$
 and $\gamma$ parameters are one by one determined by minimizing the RMSE,
 which evaluated by eq.(16). These optimized non-bonded parameters are summarized in TABLE 3.
\begin{table}[h]\centering
\small
  \caption{ Parameters of non-bonded potential represented by eqs.(3)-(5)}
  \label{tbl:example}
  \begin{tabular*}{0.9\textwidth}{@{\extracolsep{\fill}}ccccccc}
    \hline
    $\sigma_{PW}$  & $\sigma_{PA}$ & $\sigma_{PB}$ &  $\epsilon_{PW}$ & $\epsilon_{PA}$ & $\epsilon_{PB}$ 
  & $\gamma$  \\ 
    \hline
   0.40 & 0.45 & 0.46 & 2.650 & 4.356 & 3.523 & 1.13 \\
    \hline 
  \end{tabular*}
\end{table}

\subsection{Diffusion coefficients and shear viscosities of EGO/Water binary systems}
From eq.(11), the self-diffusion coefficient is obtained by evaluating of the slope of the
 mean square displacement (MSD) of the center of mass of the EGO/water molecules. The MSDs
 were calculated for each EGO and water molecules in the simulation box using 3 ns (for
 water, aqueous EGO2 and EGO4 solutions) or 30 ns (for aqueous EGO13 solution) MD-simulation.
 Dropping the first 1 ns MSD data as relaxation time, the slope of MSD versus time was
 evaluated by linear regression.  
\\
 Figure 4a and 4b show the comparison between experimental and calculated self-diffusion
 coefficients. Figure 4a shows the self-diffusion coefficients of the EGO molecules plotted
 against the EGO weight fraction ($W$) of the EGO/water binary mixtures. Figure 4b shows
 the self-diffusion coefficients of the water molecules plotted in the same manner as Figure 4a.
 Three different EGOs, which are EGO2, EGO4 and EGO13, and three different EGO weight
 fractions ($W$= 0.2, 0.5 and 0.8) are presented, including the data of aqueous EGO13
 solutions ($W$ = 0.5 and 0.8), which did not used in the parametrization of non-bonded
 potentials at Section 4.2. 
\\
Experimental observations show that the self-diffusion coefficients (both of the EGO and the
 water) are linear on a log scale when plotted against $W$. The CG-MD simulation results
 are also linear as shown in Figure 4a and 4b.
 The dependence of the $D$ on $W$ is increasing against the EGO molecular weight ($M_{W}$)
 as shown in Figure 4a, but the self-diffusion coefficients of the water molecule does
 not show such tendency (Figure 4b). The experimental $W$ and $M_{W}$-dependences of $D$ are
 reproduced by the CG-MD simulations correctly. Additionally, we performed CG-MD simulation
 of EGO45(PEG2000)/water binary mixture ($W$ = 0.2). We found that the calculated
 self-diffusion coefficients of EGO45 and water are comparable to the experimental values
 based on NMR \cite{Vergara1999} (data not shown), though these NMR measurements were observed
 at 298 K, differ from our CG-MD condition T = 293 K.
\begin{figure}[!h]
\centering
  \includegraphics[height=11cm]{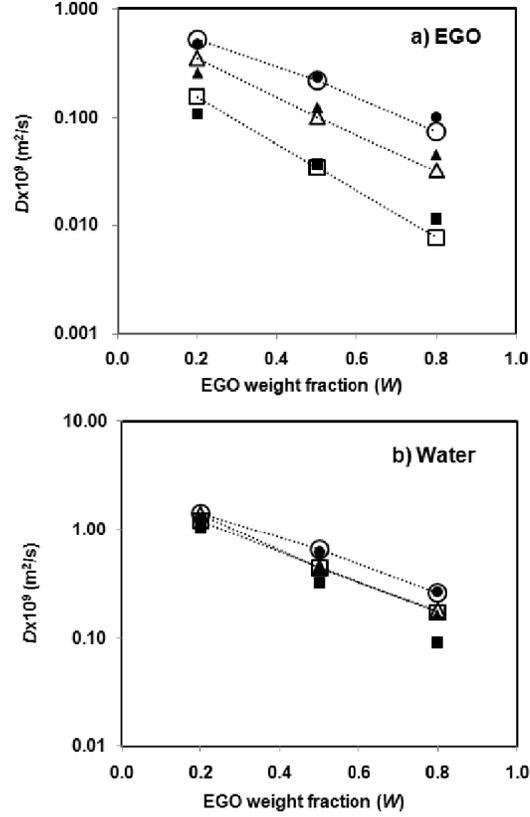}
  \caption{Self-diffusion coefficients $D$ of EGO (a) and water (b) molecules as a
 function of the weight fraction $W$ of EGO in the EGO/water binary mixtures.
 The open circles, the open triangles and the open squares denote the $D^{\rm{EXP}}$
 values measured using NMR for EGO2/water, EGO4/water and EGO13/water binary mixtures,
 respectively. The filled circles, the filled triangles and filled squares denote
 the $D^{\rm{CG}}$ values calculated using CG-MD simulations for EGO2/water,
 EGO4/water and EGO13/water binary mixtures, respectively.}
  \label{fgr:example}
\end{figure}
\\
\\
 The shear viscosity was calculated by the nonequilibrium method described by B.
 Hess \cite{Hess2002}, which estimates the shear viscosity of liquid from nonequilibrium
 simulation with an external cosine acceleration profile of the form
\begin{eqnarray}
{a_{\rm{x}}}\left({\rm{z}}\right) &=& A{\rm{cos}}\left( {k{\rm{z}}} \right),  \\
k &=&\frac{2\pi}{l_z}, \nonumber 
\end{eqnarray}
where $a_{\rm{x}}$ is the acceleration in the x direction, $l_{\rm{z}}$ is the height
 of the simulation box, $A$ is the magnitude of acceleration, z is the coordinate
 z-direction. The shear viscosity of liquid $\eta$ can be estimated by  
\begin{equation}
{\eta^{-1}}{\left(t\right)}{=}{\frac{2k^{2}}{{\rho}A}}
{\sum_{i=1}^{N}}{m_i}{v_{i,{\rm{x}}}}
{\left({t}\right)}{\rm{cos}}{\left(k{\rm{z}}_{i}\left({t}\right)\right)}/{\sum_{i=1}^{N}}{m_i},
\end{equation}
where $\rho$ is the density of a system, $m_i$ is the mass of the particle of index
 $i$, $v_{i,{\rm{x}}}$ is the velocity in the x direction of the particle of index
 $i$, $N$ is the total number of particles in the simulation box, $\eta^{-1}$ is
 the reciprocal of viscosity of the system. The $A$ parameter must be carefully
 selected, the shear rate should not be so high that the system is driven too far from
 equilibrium. The maximum share rate of the CG system $\rm{sh}_{max}^{CG}$ is
\begin{equation} 
{\rm{sh}_{max}^{CG}}=A{\frac{\rho}{\eta^{\rm{CG}}}}{\frac{l_{\rm{z}}}{2\pi}} .
\end{equation}
 For our simulations with: ${\eta^{\rm{CG}}}\approx $0.01 $[kgm^{-1}s^{-1}]$,
 $l_z\approx$ 8 [nm], $\rho\approx$1000 [kg$\rm{m}^{-3}$], and $\rm{sh}_{max}^{CG}$
 is approximately 0.2 [ps $\rm{nm}^{-1}$] $A$. This shear rate should be smaller
 than one over the longest correlation time in the system.
 For usually liquids, it will be the rotational
 correlation time of the largest molecule in the system. In the aqueous EGO13 solution
($W$ = 0.8), the rotational relaxation time of end-to-end vector of the coarse-grained
 EGO13 is approximately 6000 ps. In this case, parameter $A$ should be smaller than 0.001
 [nm $\rm{ps}^{-2}$]. When the shear rate is too high, the observed shear viscosity will
 be too low. In this article, we used $A$ = 0.0005 [nm $\rm{ps}^{-2}$] for all
 nonequilibrium CG-MD simulations. $\eta^{\rm{AA}}$ is estimated by eq.(9), with
 $S$ = 6.19 and $\eta^{\rm{CG}}$ obtained from eq.(18).
 Figure 5 shows the comparison between experimental and calculated shear viscosities
 of the aqueous EGO solutions. For nine EGO/water binary mixtures evaluated in this
 article, the symbols are same as in Fig.4, the calculated shear viscosities are
 agreed with the experiments. This means that the calculated values for two
 EGO-concentrations($W$ = 0.2 and 0.5) are very close to the experimental data and
 overall data included $W$ = 0.8 shows the same tendency as the experimental results.
 In order to verify our CG model at larger molecular weight of EGO, we performed
 the additional calculations of the EGO22(PEG1000)/water and EGO45(PEG2000)/water
 binary mixtures. At the highest EGO-concentrations ($W$ = 0.5 for EGO22 and $W$ =
 0.4 for EGO45) considered in this article, the rotational relaxation times of
 end-to-end vector of both the coarse-grained EGO22 and EGO45 are lower than the ones
 of the coarse-grained EGO13 in the aqueous EGO13 solution ($W$ = 0.8). Therefore,
 the parameter $A$ in the eq.(18) and the sampling time length of $\eta^{-1}$ are
 same as the case of the EGO13/water binary mixture ($W$ = 0.8 ). 
 Figure 6 shows the comparison between our calculated results and the experimental
 observations measured at 293 K \cite{PEG293K}. We found that the calculated viscosities are
 comparable to the literature values for aqueous EGO22 solutions. Although our
 calculated viscosities are slightly lower than the literature values for aqueous EGO45
 solutions, the tendency of the dependence of viscosity on the EGO45 concentration seems
 to be in accordance with the experiments.
\begin{figure}[!h]
\centering
  \includegraphics[height=6cm]{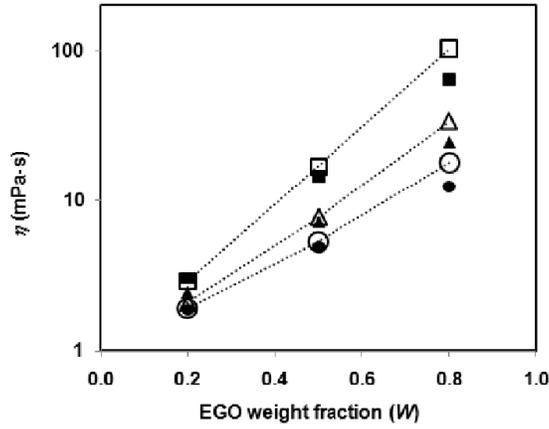}
  \caption{Shear viscosity $\eta$ of the  EGO and water binary mixture as a function
 of the EGO weight fraction $W$. The symbols are same as in Fig.4.}
  \label{fgr:example}
\end{figure}
\begin{figure}[!h]
\centering
  \includegraphics[height=6.2cm]{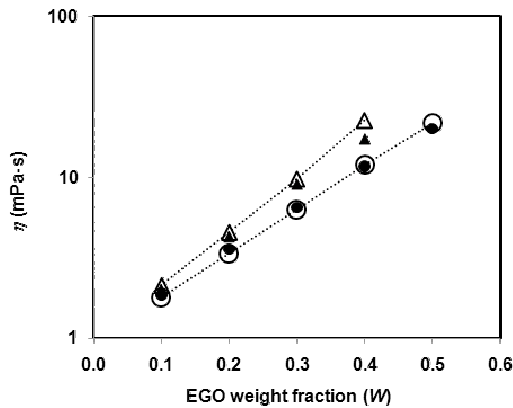}
  \caption{Shear viscosity $\eta$ of the  EGO and water binary mixture as a function
 of the EGO weight fraction $W$. The open circles and the open triangles denote
 the $\eta^{\rm{EXP}}$ values in literature \cite{PEG293K} for EGO22(PEG1000)/water
 and EGO45(PEG2000)/water binary mixtures, respectively. The filled circles and
 filled triangles denote the $\eta^{\rm{CG}}$ values calculated using
 CG-MD simulations for EGO22/water and EGO45/water binary mixtures, respectively.}
  \label{fgr:example}
\end{figure}
\newpage
\section{Conclusions}
CG-MD simulations for the EGO/water binary mixtures were performed 
at 293 K and 1 atm pressure. The EGO chain was modeled by two types 
of particles, PA and PB, which represent the oxyethylene units of
 both ends and the middle of the chain, respectively. Also, three
 water molecules are included into a single PW particle. With our
 CG model, the number of particles that should be considered in the
 simulation can be reduced ten-fold, and the time step for integration
 of Newton's equation increases ten times compared with atomistic simulations.
\\
 The parameters for the CG model were determined by the systematic
 manner, as was shown in this article. The CG bonded potential
 parameters for the EGO chain were obtained by the Boltzmann inversion
 of the corresponding atomistic distribution functions. Due to the soft
 pair potential among the CG particles, the time-scale of CG-MD
 simulation is not equivalent to a realistic time scale. In order to
 estimate the proper dynamical properties by means of CG-MD
 simulations, the scaling relation for time in the CG model is
 introduced. Moreover, the hydrodynamic radius of PW particles in
 our CG model is larger than those of atomistic water molecules,
 due to the gain in volume of the PW particle as the result of the
 coarse graining of water, as shown in Figure 1. Therefore, we also
 introduced the scaling relation for the water diffusivity based on
 the Stokes-Einstein law.
\\
 With the bonded potential parameters and the scaling relations,
 the 12-6 Lennard-Jones non-bonded potential parameters are
 straight-forwardly determined to reproduce the experimental
 observations (the density and the self-diffusion coefficient).
 We adopted the Lorentz-Berthelot mixing rule for the non-bonded
 interactions between the unlike particles. In this article, the
 Lorentz-Berthelot mixing rule is slightly modified, as shown in
 eq.(4) and eq.(5), to estimate the proper miscibility of the EGO13
 in water. By using the determined CG force-field parameters for
 the EGO/water binary mixtures, our CG-MD simulation gives the
 estimations which agreed well with the experimental shear-viscosity data,
 including of the PEG1000/water and the PEG2000/water binary mixtures
 which were not used in our parameterization procedure.  
 The largest simulation in this article corresponds to a
 1.2 $\mu$s atomistic simulation for 100,000 atoms. 
 Our CG model with the parameterization scheme for the CG particles
 may be useful to study of the dynamic properties of a liquid which
 contains relatively low molecular weight polymers or oligomers.
\\
 In our future work, we plan to investigate the CG models for the several
 watersoluble polymers (e.g., polypropylene glycol, polyvinyl pyrrolidone,
 polyvinyl alcohol), for the estimations of the shear-viscosity and
 the diffusivity of these water mixtures through CG-MD simulations.



%% References


%%\bibliographystyle{model1a-num-names}
%%\bibliography{<your-bib-database>}






% \end{thebibliography}

\end{document}